\documentclass[%
 reprint, 
 amsmath,amssymb,
 aps,prl,
]{revtex4-1}

\usepackage{graphicx}
\usepackage{dcolumn}
\usepackage{bm}
\usepackage{siunitx}
\usepackage[caption=false]{subfig}
\usepackage[english]{babel}
\begin{document}

\title{Phase-field collagen fibrils: Coupling chirality and density modulations}
\author{Samuel Cameron}
\author{Laurent Kreplak}
\author{Andrew D Rutenberg}
 \email{adr@dal.ca}
\affiliation{
Dept. of Physics and Atmospheric Science, Dalhousie University, Halifax NS, Canada}%
\date{\today} 

\begin{abstract}
To describe the interaction between longitudinal density modulations along collagen fibrils (the D-band) with the radial twist-field of molecular orientation (double-twist), we couple phase-field-crystal (PFC) with liquid-crystalline free-energies to obtain a hybrid model of equilibrium collagen fibril structure.  We numerically compute the resulting axial and radial structure. We find two distinct fibrillar phases, `L' and `C', with a coexistence line that ends in an Ising-like critical point. We propose that coexistence between these phases can explain the bimodal distribution of fibril radii that has been widely reported within tendon tissues. Tensile strain applied to our model fibrils straightens the average fibrillar twist and flattens the D-band modulation. Our PFC approach should apply directly to other longitudinally-modulated chiral filaments, such as fibrin and intermediate filaments.
\end{abstract}
\maketitle

\textit{Introduction} --- Collagen molecules assemble into cylindrical fibrils, which exhibit a wide range of possible radii, $R\in10-200\si{\nano\meter}$ depending on anatomical location in vivo \cite{Parry:1978bb, Raspanti:2018} and self-assembly conditions in vitro \cite{Gobeaux:2008, Harris:2013dv, Harris:2007iw, Asgari:2017}. Collagen molecules are chiral \cite{Rich:1955vc}, and collagen fibrils are chiral materials. Collagen fibrils serve as building blocks in mechanically-loaded tissues such as tendon, skin, and bone \cite{Sherman:2015ei}. The crucial role of fibrils within the human body highlights the importance of understanding the interplay of their chiral and mechanical properties.

The chiral nature of collagen fibrils is evident in the tilted alignment of individual molecules with respect to the fibril axis  \cite{Raspanti:interjbiomacromol1989, Ottani:2001, Brown:softmatter2014, Hulmes:biophys1995, Cameron:2018kq}. This ``twist'' angle $\psi$  can be as large as $\SI{17}{\degree}$ at the surface of corneal fibrils, and is approximately $\SI{5}{\degree}$ in tendon fibrils \cite{Brodsky:1980,Raspanti:2018}.   Theoretical work  treating the fibril structure as a chiral liquid crystal \cite{Brown:softmatter2014, Cameron:2018kq, Grason:2007bs} has shown that twisted fibrils can be thermodynamically stable, and predicts that twist continuously varies within the fibril as a ``double-twist'' field $\psi(r)$.

Periodic density modulations along the fibril axis are ubiquitous. These modulations originate from specific intermolecular interactions, as described in the Hodge-Petruska model \cite{petruska:1964in}. The most prominent density modulation is known as the D-band. The D-band period is remarkably consistent for ex vivo fibrils, between $64-67\si{\nano\meter}$ \cite{Fang:2013gb}. D-band periods are somewhat more variable for in vitro fibrils \cite{Fang:2013ba} and can be manipulated by the mixture of different collagen types within the fibril \cite{Asgari:2017}. The ``gap'' and ``overlap'' regions of the Hodge-Petruska model suggest a D-band modulation amplitude that is $10\%$ of the total density. However, a simple geometrical interpretation of the Hodge-Petruska model also locally implies a D-band period $d \propto \cos\psi(r)$ within an individual fibril \cite{Raspanti:2018}, whereas only a single D-band period is observed in experiment. One previous model of both D-band and double-twist therefore assumes a radially constant twist \cite{Galloway:1985is}, though this is energetically unfavourable for the double-twist field \cite{Cameron:2018kq}. A second model has an approximately constant twist gradient \cite{Raspanti:interjbiomacromol1989}, which is energetically preferable for the double-twist but implies local stretching or compression of the D-band \cite{Raspanti:2018}. These two models of the radial twist $\psi(r)$ represent the opposite limits of a stiff or soft D-band, respectively. How can we explore the coupling of the D-band and double-twist more generally? 

While atomistic molecular dynamics (MD) simulations of the D-band, e.g. \cite{Gautieri:nanolett2011}, can explore axial strain \cite{Varma:2016dq}, they currently ignore radial twist. This is because a continually varying $\psi(r)$ precludes unit cells with small numbers of molecules.  Fortunately, coarse-grained phase-field-crystal (PFC) approaches for addressing periodic modulations of crystalline materials \cite{Elder:2004ct} are compatible with coarse-grained models of the radial twist. PFC models impose density modulations with coarse-grained fields, and so do not require the inefficiently-short atomic time-scales of MD. PFC models allow us to quantitatively explore both mechanical and structural properties of the fibril for arbitrary values of the D-band stiffness. 

\textit{Phase-field-crystal fibril} --- PFC theory adds terms to the coarse-grained free energy to generate periodic structure. PFC is particularly simple in one dimension, such as along the collagen fibril. From the Hodge-Petruska model, molecules pack along their long axis with period $\tilde{d}_{\parallel}=\SI{67}{\nano\meter}$ in the absence of molecular twist.  We can write the PFC contribution to the free-energy per unit volume as
\begin{align}\label{eq:Epfc}
    \tilde{E}_{\mathrm{pfc}}=&\frac{1}{\pi \tilde{R}^2 \tilde{L}} \tilde{\Lambda} \int d^3\tilde{x} {\tilde{\phi}}(\tilde{\bm{r}})\bigg(\frac{4\pi^2}{\tilde{d}_{\parallel}^2}+\tilde{\nabla}_{\parallel}^2\bigg)^2{\tilde{\phi}}(\tilde{\bm{r}})\nonumber\\
    +& \frac{1}{\pi \tilde{R}^2 \tilde{L}} \tilde{\omega} \int d^3 \tilde{x} {\tilde{\phi}}^2({\tilde{\phi}}^2-{\tilde{\chi}}^2),
\end{align}  
where we integrate over a cylindrical fibril of radius $\tilde{R}$ and length $\tilde{L}$, and $\tilde{\phi}$ is the amplitude of modulations due to the D-band. $\tilde{\nabla}_{\parallel}$ is the gradient operator in the direction parallel to the local molecular orientation. The first integral of eqn \ref{eq:Epfc} is minimized when the modulations have the same local periodicity as $\tilde{d}_{\parallel}$, and $\tilde{\Lambda}$ characterizes the D-band stiffness. For  $\tilde{\chi}^2>0$, the second integral is minimized when $\tilde{\phi}^2$ is non-zero -- which determines a preferred non-zero D-band amplitude. $\tilde{\omega}$ characterizes the energetics of D-band formation.

We can further simplify these PFC contributions. Under a single-mode approximation for the D-band, we take  $\tilde{\phi}(\tilde{z}) = \tilde{\delta}\cos(\tilde{\eta}\tilde{z})$ -- where $\tilde{\eta}$ is the observed D-band wavenumber.  Furthermore, we  work within the ansatz that the molecular orientation is determined by the twist-field: $\bm{n}=\sin\psi(\tilde{r})\hat{\phi}+\cos\psi(\tilde{r})\hat{z}$ \cite{Brown:softmatter2014, Cameron:2018kq}. Because the local orientation is not along the fibril axis, this couples the D-band periodicity to $\psi(r)$ through the gradient in the frame parallel to $\bm{n}$, 
$\tilde{\nabla}_{\parallel}=\cos\psi(\tilde{r})\partial/\partial \tilde{z}$. We then obtain a simpler expression
\begin{align}\label{eq:EpfcSimpler}
    \tilde{E}_{\mathrm{pfc}}=&
        \frac{\tilde{\Lambda}\tilde{\delta}^2}{2\tilde{R}^2}\int_0^{\tilde{R}}\tilde{r}d\tilde{r}\bigg(\frac{4\pi^2}{\tilde{d}_{\parallel}^2}-\tilde{\eta}^2\cos^2\psi(\tilde{r})\bigg)^2\nonumber\\
        &+\frac{\tilde{\omega}{\tilde{\delta}}^2}{2}\bigg(\frac{3}{4}{\tilde{\delta}}^2-{\tilde{\chi}}^2\bigg).
\end{align}

To this we add the volume averaged Frank free energy density for the double-twist director field (see \cite{Cameron:2018kq}):
\begin{align}\label{eq:Efrank}
    \tilde{E}_{\mathrm{Frank}}=&\frac{2}{\tilde{R}^2}\int_0^{\tilde{R}}\tilde{r}d\tilde{r}\bigg(\frac{1}{2}\tilde{K}_{22}\bigg(\tilde{q}-\frac{\partial\psi}{\partial\tilde{r}}-\frac{\sin2\psi}{2\tilde{r}}\bigg)^2\nonumber\\
&+\frac{1}{2}\tilde{K}_{33}\frac{\sin^4\psi}{\tilde{r}^2}\bigg)-(\tilde{K}_{22}+\tilde{k}_{24})\frac{\sin^2\psi(\tilde{R})}{\tilde{R}^2},
\end{align}
where $\tilde{K}_{22}$, $\tilde{K}_{33}$, and $\tilde{k}_{24}$ are the usual Frank elastic constants (twist, bend, and saddle-splay, respectively). We also add the average surface energy per unit volume due to free interfaces, given by $\tilde{E}_{\mathrm{surf}}=2\tilde{\gamma}/\tilde{R}$, where $\tilde{\gamma}$ is the surface tension.  

For the remainder of this paper, variables without a tilde will be dimensionless. We do this by measuring energies in units of $\tilde{K}_{22}\tilde{q}^2$, measuring density in units of $\tilde{\chi}$, measuring radius in units of $1/\tilde{q}$, and the D-band wavenumber in units of $1/\tilde{d}_{\parallel}$. Combining $E_{\mathrm{pfc}}$, $E_{\mathrm{Frank}}$, and $E_{\mathrm{surf}}$, we obtain the total average free energy density of the fibril as a function of radius $R$, D-band modulation amplitude $\delta$, D-band modulation period $2\pi/\eta$, and twist angle field $\psi(r)$: $E_{\mathrm{tot}}= E_{\mathrm{pfc}}+ E_{\mathrm{Frank}}+E_{\mathrm{surf}}$.

There are five dimensionless parameters that control the behaviour of our system. $K_{33} \equiv \tilde{K}_{33}/\tilde{K}_{22}$ and $k_{24} \equiv \tilde{k}_{24}/\tilde{K}_{22}$ characterize the bend and saddle-splay elastic constants. Consistent with our previous work we will fix $K_{33}=30$ \cite{Cameron:2018kq, Lee:liqcryst1990}.   $\gamma \equiv \tilde{\gamma}/(\tilde{K}_{22}\tilde{q})$ controls the surface tension. $\Lambda \equiv 2\tilde{\Lambda}\tilde{\chi}^2/(3\tilde{K}_{22}\tilde{q}^2\tilde{d}_{\parallel}^4)$ controls the coupling strength between the D-band and the molecular twist and can be related to the Young's modulus at zero twist.  $\omega \equiv 2\tilde{\omega}\tilde{\chi}^4/(3\tilde{K}_{22}\tilde{q}^2)$ controls the strength of the D-band double well potential, which cannot be any larger than the polymerization energy of collagen fibrils \cite{Kadler:1987}.

We minimize $E_{\mathrm{tot}}$ with respect to $\psi(r)$ for chosen initial values of $R$, $\eta$, and $\delta$ using a numerical implementation of the corresponding Euler-Lagrange equations \cite{Brown:softmatter2014}, and obtain $\psi_0(r)$.  We then minimize $E_0 \equiv E(R,\eta,\delta;\psi_0(r))$ with respect to $R$ to obtain a global (thermodynamic) minimization of $E_{\mathrm{tot}}$. Our numerical minimization routines are available online via GitHub \cite{github-repoSam}.  The default parameter values, which apply unless otherwise stated, are $\gamma=0.04$, $k_{24}=0.5$, $\tilde{q}=\SI{4}{\per\micro\meter}$, $\Lambda=600$, and $\omega=20$ --- and are discussed below.

\begin{figure*}[t!]
\hspace{-0.8cm}
    \subfloat{\includegraphics[width=2.5in]{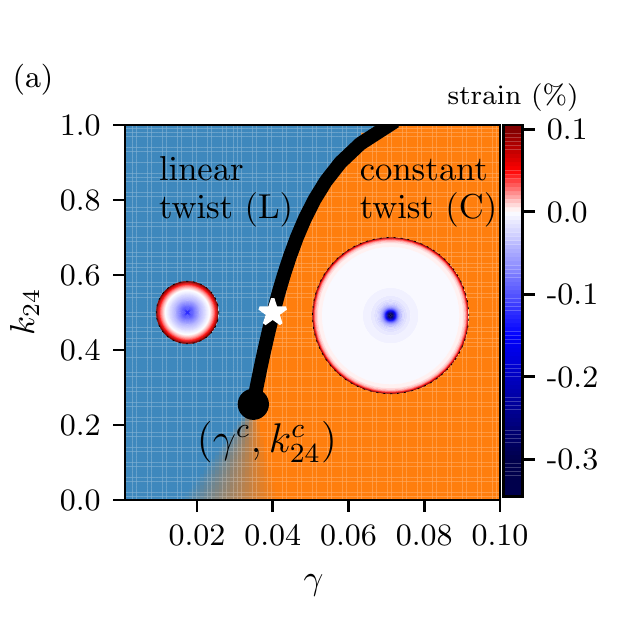}\label{fig:1a}}\hspace{0cm}
    \subfloat{\includegraphics[width=2.5in]{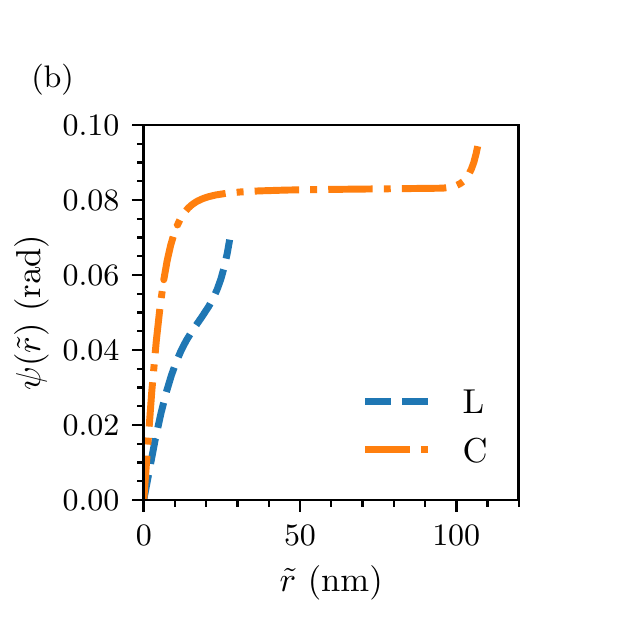}\label{fig:1b}}\hspace{-0.8cm}
    \subfloat{\includegraphics[width=2.2in]{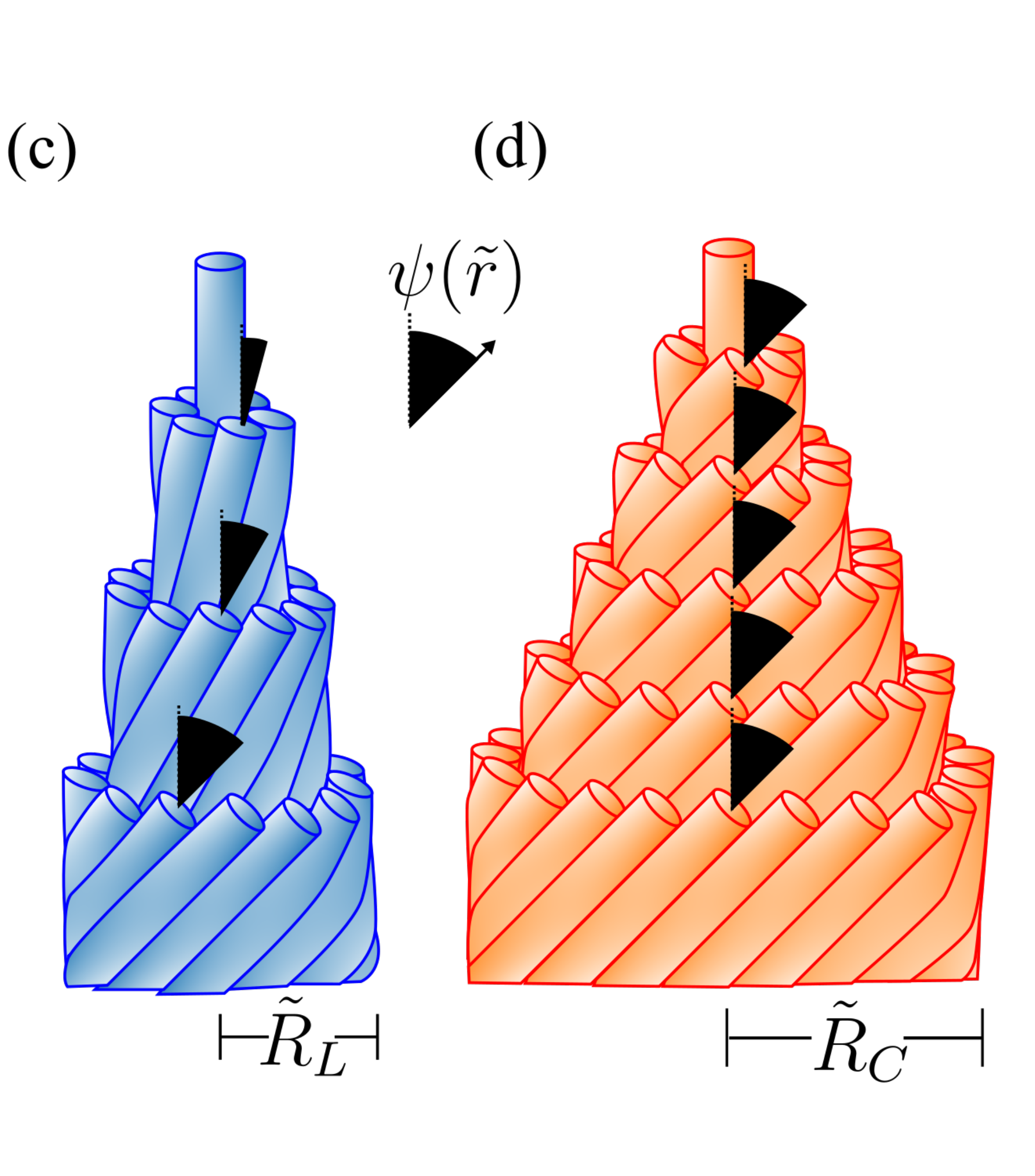}\label{fig:1c}}\hspace{0.5cm}
    \subfloat{\label{fig:1d}}
    \caption{\protect\subref{fig:1a} Phase diagram of the PFC fibril model in the dimensionless surface-tension vs saddle-splay elastic constant ($\gamma$, $k_{24}$) plane. A linear twist (L, indicated by darker blue shading)  phase and constant twist (C, indicated by lighter orange shading) phase are separated by a coexistence line (thick black line), which ends at the critical point $\gamma^c=0.0348$, $k_{24}^c=0.2554$ (black circle). Inset fibril cross-sections show the equilibrium axial strain that maintains the constant D-band period $2\pi/\eta$ at ($\gamma=0.04$, $k_{24}=0.5$) -- indicated by the white star. The local compression and tension are indicated by blue and red shading, respectively, as indicated by the scale bar, while regions with no local strain are white. \protect\subref{fig:1b} $\psi(r)$ for the coexisting linear (blue, dashed) and constant (orange, dot-dashed) twist phases of the PFC fibril model corresponding to the cross-sections in \protect\subref{fig:1a}. The constant twist phase has $\psi(r)\simeq\psi_0$ throughout most of the fibril (here $\psi_0 \approx \SI{0.08}{\radian}$). Schematics of the linear and constant twist fibril phases are shown in \protect\subref{fig:1c} and \protect\subref{fig:1d}, respectively. Note that $\psi(0)=0$ from energetic considerations, and that the fibril radius in the C-phase  exceeds that in the L-phase ($R_C> R_L$).}\label{fig:1}
\end{figure*}

\textit{Coexistence} --- In Fig.~\ref{fig:1a} we show the phase diagram of the PFC fibril model in the $\gamma$, $k_{24}$ plane. The D-band modulation leads to a co-existence line (indicated by the thick black line) between qualitatively distinct fibril phases, which ends at a critical point at $\gamma^c=0.0348$, $k_{24}^c=0.2554$.  This coexistence line, along which the fibril radius and twist-field changes discontinuously, is not observed without the PFC terms \cite{Cameron:2018kq}. 

For the default parameterization, indicated by a white star in Fig.~\ref{fig:1a}, we show the two coexisting twist-field solutions which minimize $E_{\mathrm{tot}}$ in Fig.~\ref{fig:1b}. For smaller $R$ (blue line), the equilibrium twist-field $\psi(r)$ has an approximately constant gradient and we call it the ``linear twist'' phase (L). For larger $R$ (orange line), $\psi(r)$ has a large region of approximately constant twist and we call it the ``constant twist'' phase (C). From a molecular twist perspective, the C phase can be viewed as a ``core-shell'' structure with the core showing a strong linear twist gradient while the shell shows a constant twist -- albeit with a narrow region of additional twist gradient at the surface. 

Qualitatively, fibrils are energetically stabilized by surface twist through $k_{24}$ as well as by the D-band amplitude $\delta$. Larger values of surface tension $\gamma$ drive larger $R$, to reduce the surface area per unit volume. These larger radii also lead to increasingly large D-band elastic energy through $\Lambda$, which can then be reduced by a region of constant twist. 

The L and C twist-fields are visualized schematically in Fig.~\ref{fig:1c} and \ref{fig:1d}, respectively. The inset circles within Fig.~\ref{fig:1a} indicate the axial D-band strain in a cross-section of unstretched fibrils corresponding to the blue (L) and orange (C) curves of Fig.~\ref{fig:1b} (with D-band periodicity $d_L/d_{||}=2\pi/\eta_L=0.999$ and $d_C/d_{||}=2\pi/\eta_C=0.997$), respectively. For the L phase, most of the fibril is strained ranging from the centre under axial compression of $0.3\%$ to the surface under tension of $0.1\%$. For the C phase, only the centre and surface are significantly strained. 

\begin{figure}[t]
    \includegraphics[width=2.8in]{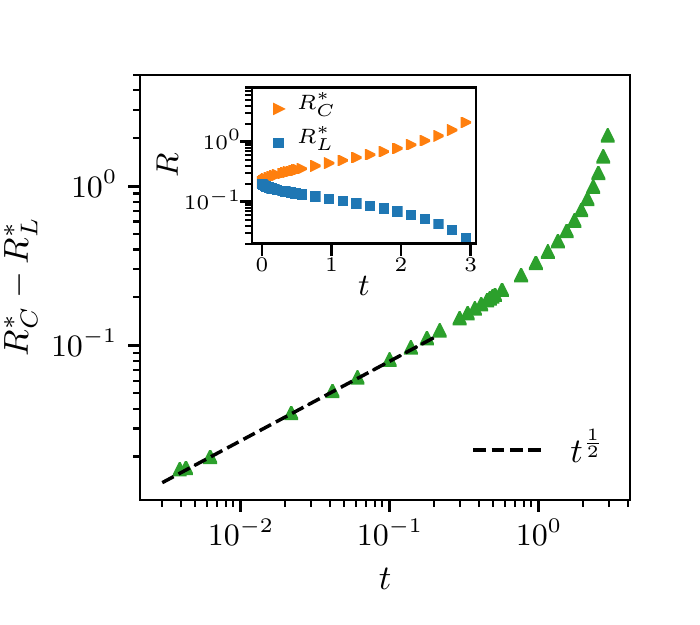}\vspace{-0.5cm}
    \caption{The difference of fibril radii of the linear twist fibril phase, $R_L^*$ and constant twist fibril phase, $R_C^*$, (green triangles) vs distance $t \equiv k_{24}/k_{24}^c-1$ along the coexistence line shown in Fig.~\protect\ref{fig:1}. The inset shows the separate radii, as indicated. Near the critical point ($\gamma^c$, $k_{24}^c$) the discontinuity across the coexistence line vanishes as $t^{1/2}$ -- as indicated by the dashed black line.}    \label{fig:2}
\end{figure}

Our coarse-grained free-energy approach does not include any fluctuations, so we expect mean-field critical exponents \cite{Chaikin:318158} to describe the discontinuities across the coexistence line sufficiently close to the critical point. In Fig.~\ref{fig:2}, we show the difference in radii of the coexisting linear and constant twist phases, $R_L^*$ and $R_C^*$ respectively, vs the distance from the critical point $t \equiv k_{24}/k_{24}^c-1$. We find an Ising-like mean-field critical exponent $\beta =1/2$, as indicated by the dashed black line. In the inset, we show the splitting of $R$ into $R_L^*$ and $R_C^*$ near the critical point. Near $t\simeq3$ we see that the ratio $R_C^*/R_L^*$ can be as large as $100$.  

Remarkably, the coexistence of widely different fibril radii have been reported \cite{Goh:2012} (see also \cite{PattersonKane:1997, Kalson:2015}) within tendon samples. The ratio of radii increases with age, and approaches 5 for older tendon \cite{Goh:2012}. We propose that tendon fibrils are in coexistence within our equilibrium model. Indeed, our default parameters are chosen to approximately recapitulate tendon fibril properties.  $\gamma=0.04$ and $k_{24}=0.5$ on the coexistence line  are chosen to recover $R_C^*/R_L^*\simeq 4$. Taking $\tilde{q}=\SI{4}{\per\micro\meter}$ is consistent with in vitro studies of chiral nematic collagen solutions \cite{DeSaPeixoto:2011jm}, and leads to radii $\tilde{R}_C \approx \SI{106}{\nano\meter}$ and $\tilde{R}_L \approx \SI{27}{\nano\meter}$.

\textit{Elastic properties} --- We can probe the tensile response of our model fibril to an applied strain $\epsilon$. To do this, we axially strain the fibril and D-band while conserving volume by imposing $\eta=\eta_{\mathrm{eq}}/(1+\epsilon)$ and $R=R_{\mathrm{eq}}/\sqrt{1+\epsilon}$, where $\eta_{\mathrm{eq}}$ and $R_{\mathrm{eq}}$ are the unstrained values. We then minimize $E_{\mathrm{tot}}$ with respect to $\delta$ and $\psi(r)$ at these strained values of $\eta$ and $R$.

\begin{figure}[t]
    \subfloat{\includegraphics[width=3.1in]{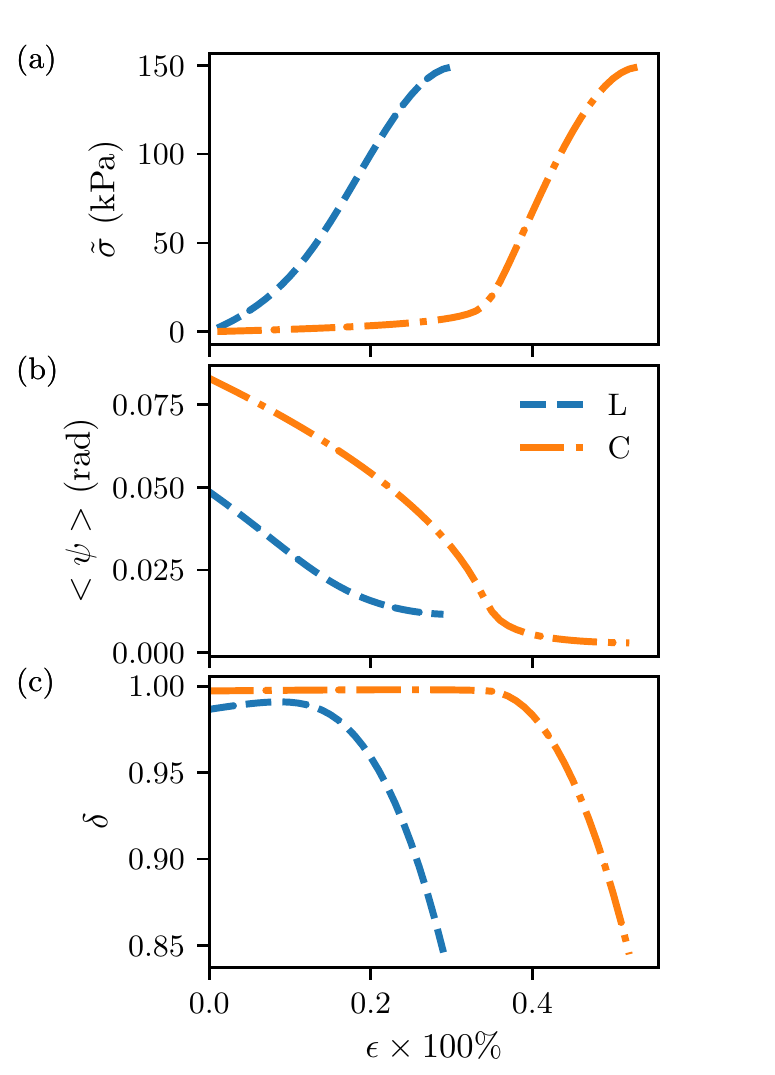}\label{fig:3a}}
    \subfloat{\label{fig:3b}}
    \subfloat{\label{fig:3c}}
	\caption{Mechanical and structural properties of linear twist (blue dashed) and constant twist (orange dash dotted) tendon fibrils. \protect\subref{fig:3a} Stress vs strain, \protect\subref{fig:3b} volume average tilt $\langle \psi \rangle$ vs strain, \protect\subref{fig:3c} D-band amplitude $\delta$ vs strain.}\label{fig:3}
\end{figure}

Fig.~\ref{fig:3} shows the resulting \protect\subref{fig:3a} stress $\tilde{\sigma}=dE/d\epsilon$, \protect\subref{fig:3b} D-band amplitude $\delta$, and \protect\subref{fig:3c} volume average twist $\langle \psi(r) \rangle$ vs strain $\epsilon$.  As $\epsilon$ increases, the fibrils stiffen until they reach a maximum stress -- after which the fibrils are unstable. This stiffening is delayed in strain and dramatically larger for the constant twist fibrils, giving rise to two approximately linear regimes of stress vs strain. For the linear twist fibrils, only the second regime is observed. 

Untwisting (decreasing $\langle \psi \rangle$, as seen in Fig.~\ref{fig:3}b)) and flattening of the D-band modulation (as seen in Fig.~\ref{fig:3}) occur as strain increases.  The untwisting and flattening with strain result from the coupling between the D-band and the radial-twist in our model. 

By choosing $\Lambda=600$ we effectively determine $\tilde{Y}_{high} \approx \SI{100}{\mega\pascal}$ to be comparable with the maximal slope of $\tilde{Y} \simeq \SI{30}{\mega\pascal}$ observed in non-cross-linked fibrils \cite{Graham:2004bl}, using $\tilde{K}_{22}=\SI{6}{\pico\newton}$ \cite{Cameron:2018kq}.   However, the observation of a significant low-slope regime is also observed experimentally \cite{Graham:2004bl} with a comparable $\tilde{Y}_{low} \simeq \SI{1}-\SI{5}{\mega\pascal}$. Only $\omega$ is relatively unconstrained, but that is because (data not shown) our results are qualitatively independent of it.  We take $\omega=20$.

\textit{Discussion} --- The phase coexistence (Fig.~\ref{fig:1}), that emerges from the interaction between the axial D-band modulation and the radial double-twist, results in strikingly different fibril radii coexisting in thermodynamic equilibrium (Fig.~\ref{fig:2}) and provides a natural explanation for experimentally observed bimodal radius distributions of tendon fibrils \cite{Goh:2012, PattersonKane:1997, Kalson:2015}. 

Coupling the D-band and twist-fields also leads us to predict that elastic strain straightens the twist-field (Fig.~\ref{fig:3b}) and flattens the D-band amplitude (Fig.~\ref{fig:3c}). Qualitatively similar strain-straightening has been observed in recent synchrotron X-ray scattering studies of corneal fibrils \cite{Bell:2018ix}. Similarly, recent creep studies of modestly strained tendon completely degrades the D-band after several hours \cite{Hijazi:2019}, which is also consistent with our results. 

We have found that coexisting phases have distinct twist fields $\psi(r)$, with the smaller radii fibrils qualitatively similar to constant gradient models discussed by Raspanti \cite{Raspanti:2018}, while the larger fibrils are qualitatively similar to constant-tilt models proposed for corneal fibrils \cite{Galloway:1985is, Ottani:2001,Silver:1992fl}. We find that the equilibrium axial strain (Fig.~\ref{fig:1a} insets) within these fibrils is quite distinct in local variation -- though not in magnitude. Both L and C  fibrils are under considerable compression at the centre and tension at the surface. 

We also find that the twist-field has dramatic impact on the elastic response of fibrils, as seen comparing the L and C phase fibrils in Fig.~\ref{fig:3a}.  This contrast may not be observed for \emph{in vivo} fibrils where covalent cross-linking greatly stiffen fibrils \cite{Silver:2003}. Nevertheless, carefully controlled assembly conditions in vitro can achieve a wide range of fibril radii and twist-fields \cite{Cameron:2018kq}. While co-existence has not been directly reported for in vitro assembly, there is a striking 10-fold increase in fibril radius over a narrow range of  precursor concentration from $75-100$  mg/ml \cite{Gobeaux:2008}. This is consistent with the discontinuity observed along co-existence in Fig.~\ref{fig:2} and we would therefore expect significant differences in the stress-strain curves as well. Further exploration of in vitro assembled fibrils may also offer a way of exploring the critical point of Fig.~\ref{fig:1a}, and of characterizing fluctuation effects there. 

We can obtain a general relationship between the observed twist-field and D-band periodicity by minimizing the first term in eq.~\ref{eq:EpfcSimpler} with respect to $d/d_{||}= 2 \pi/\eta$:
\begin{equation} \label{eq:d}
    d_{\mathrm{eq}}/d_\parallel = \sqrt{\langle \cos^4 \psi \rangle/
                           \langle \cos^2 \psi \rangle},
\end{equation}
where this equation holds for all parameter values. Approximating the twist-field as constant, we obtain $d_C/d_{||} = \cos \psi(R)$. Approximating the twist-field as having a constant gradient, i.e. $\psi(r) = ar$, we obtain $d_L/d_{||} = 1-\psi(R)^2/4 + 19 \psi(R)^4/288+O(\psi(R)^6)$.  While we have not discussed corneal fibrils so far, $\psi(R) \simeq 0.3$, while $d_{\mathrm{cornea}}/d_{||} = 64nm/67nm \simeq 0.955$ \cite{Raspanti:2018}. This indicates that corneal fibrils are in the C phase --- confirming results from a previous electron tomography study \cite{Holmes:2001}.

While we have described the equilibrium fibril structure that minimizes $E_{\mathrm{tot}}$, we can expand around that minimum and consider equilibrium fluctuations. In particular, consider D-band periods $d=2\pi(1+u)/\eta_{\mathrm{eq}}$, where $2\pi/\eta_{\mathrm{eq}}$ is the equilibrium period. We have that $E(u)-E(0)\approx1/2\tilde{Y}u^2$. Multiplying by the volume in a single D-band period, and using the Boltzmann distribution we obtain $P(u)\propto\exp\left(-u^2/(2 \sigma^2_{d/d_{eq}}) \right)$, where
\begin{equation} 
    \sigma_{d/d_{\mathrm{eq}}}^2 = k_BT/(\pi\tilde{R}^2\tilde{d}_{\mathrm{eq}}\tilde{Y}). \label{eq:sigma}
\end{equation} 
With $k_BT\approx\SI{4.114e-3}{\pico\newton\micro\meter}$, $\tilde{R}$, and $\tilde{Y}$, we obtain $\sigma_C \simeq \sigma_L \simeq 0.001$.  AFM studies of uncrosslinked fibrils have reported a narrow, approximately Gaussian distribution of D-band spacings with a fractional width $\sigma_{d/d_\parallel} \approx 1\%$ \cite{Fang:2013ba} --- approximately tenfold larger than our prediction.  While this indicates that the coupling we have used between the D-band and twist field is plausible, it suggests that additional physics is needed to describe the observed D-band variability. One possibility is that the reported longitudinal variability of mechanical properties along collagen fibrils \cite{MinaryJolandan:2009} may lead to an increased D-band variability as well.

We have seen that using a one-dimensional phase-field crystal (PFC) approach to couple longitudinal D-band modulations with the radial twist-field leads to predict two distinct structural phases for collagen fibrils: linear twist (L) and constant twist (C). We find phase-coexistence between L and C that could describe the bimodal radius distribution of tendon collagen fibrils observed \textit{in vivo}. It should be possible to explore the critical point with \textit{in vitro} fibril assembly systems. 

This PFC approach to coupling of density modulations with orientation fields could also be applied more generally in chiral materials \cite{Grason:2015, Grason:2016}. It should also directly apply to a number of chiral self-assembling fibrillar systems that exhibit longitudinal modulations, with only the parameterization to be determined. These include fibrin \cite{Weisel:1986, Weisel:1987}, keratin filaments \cite{Aebi:1983}, and nuclear lamin paracrystals \cite{Aebi:1986}.

\bibliography{bib}
\end{document}